\documentclass[useAMS,usenatbib,usegraphicx]{mn2e}
\usepackage{amsmath,amssymb,amsbsy,amsfonts}
\usepackage[british]{babel}
\usepackage{mdwmath}
\usepackage{color}
\usepackage{url}
\usepackage{mathrsfs}
\usepackage{array}
\usepackage{graphicx}
\usepackage{units}
\usepackage{hyperref}
\usepackage{times}
\usepackage{pdflscape}

\newcommand{\eq}[1]{Eq.\,\eqref{#1}}
\newcommand{\eqs}[1]{Eqs.\,\eqref{#1}}
\newcommand{\cc}{\mathrm{c}}

\newcommand{\ud}{\mathrm{d}}
\newcommand{\G}{\mathrm{G}}
\newcommand{\s}[1]{{\text{\tiny $#1$ }}\hspace{-2pt}}
\newcommand{\rg}{r_\s{\mathrm{g}}}
\def\Msun{\mathrm{M}_{\odot} }

\renewcommand{\theequation}{\thesection.\arabic{equation}}

\title[Newtonian description of Schwarzschild spacetime dynamics]
{An accurate Newtonian description of particle motion around a Schwarzschild
black hole}
\author[E.~Tejeda \& S.~Rosswog]{Emilio Tejeda\thanks{Corresponding author:
emilio.tejeda@astro.su.se} and Stephan Rosswog \\ Department of Astronomy and 
Oskar Klein Centre, Stockholm University, AlbaNova, SE-10691 Stockholm, Sweden}

\pagerange{\pageref{firstpage}--\pageref{lastpage}}

\begin{document}

\maketitle

\label{firstpage}

\begin{abstract}
A generalized Newtonian potential is derived from the geodesic motion of test
particles in Schwarzschild spacetime. This potential reproduces several
relativistic features with higher accuracy than commonly used pseudo-Newtonian
approaches. The new potential reproduces the exact location of the marginally
stable, marginally bound, and photon circular orbits, as well as the exact
radial dependence of the binding energy and the angular momentum of these
orbits. Moreover, it reproduces the orbital and epicyclic angular frequencies to
better than $6\%$. In addition, the spatial projections of general trajectories
coincide with their relativistic counterparts, while the time evolution of
parabolic-like trajectories and the pericentre advance of elliptical-like
trajectories are both reproduced exactly. We apply this approach to a standard
thin accretion disc and find that the efficiency of energy extraction agrees to
within $3\%$ with the exact relativistic value, while the energy flux per unit
area as a function of radius is reproduced everywhere to better than $7\%$. As a
further astrophysical application we implement the new approach within a
smoothed particle hydrodynamics code and study the tidal disruption of a main
sequence star by a supermassive black hole. The results obtained are in very
good agreement with previous relativistic simulations of tidal disruptions in
Schwarzschild spacetime. The equations of motion derived from this potential can
be implemented easily within existing Newtonian hydrodynamics codes with hardly
any additional computational effort.
\end{abstract}


\section{Introduction} 

Our current understanding of some of the most energetic phenomena in the
Universe (such as active galactic nuclei, X-ray binaries and gamma-ray bursts)
involves the accretion of gas onto astrophysical black holes as the underlying
mechanism for powering these sources \citep[see e.g.][]{king02}. It is clear
that a satisfactory study of any of these systems should include a consistent
treatment of the strong gravitational fields found in the vicinity of these
relativistic objects. It is nonetheless remarkable that many of the early works
on the subject, which were essentially Newtonian with a few general relativistic
effects incorporated `by hand', proved to be very successful at modelling a
variety of accreting systems. For instance, the standard thin disc model of
\cite{shakura} is purely Newtonian and the only result from general relativity
that it uses is the existence of the marginally stable circular orbit at which
their disc model was truncated. Similarly, the model of a thick accretion disc
introduced by \cite{pw} (PW hereafter) is based on Newtonian dynamics with the
substitution of the Newtonian gravitational potential $\Phi_\s{\mathrm{N}} = -\G
M/r$ by the pseudo-Newtonian potential\footnote{The qualifier `pseudo' is used
to indicate that the related potential does not satisfy the Poisson equation
$\nabla^2\Phi= 4\pi\G\rho$.}
\begin{equation}
\Phi_\s{\mathrm{PW}}(r) = -\frac{\G M}{r-2\,\rg},
\label{e1.1} 
\end{equation}
where $\rg=\G M/\cc^2$ is the so-called gravitational radius. The potential
$\Phi_\s{\mathrm{PW}}$ not only reproduces the correct location of the
marginally stable and marginally bound circular orbits around a Schwarzschild
black hole (at $r_\s{\mathrm{ms}} = 6\,\rg$ and $r_\s{\mathrm{mb}} = 4\,\rg$,
respectively), but it also gives reasonable approximations to other quantities,
e.g.~binding energy (percentage error (p.e.)~$\leqslant 13\%$) and angular
momentum of circular orbits (p.e.~$\leqslant 6\%$). Nevertheless, other
quantities such as the orbital frequency $\Omega$ and the epicyclic frequency
$\Omega^\parallel$ are not accurately reproduced (p.e.~$\leqslant 50\%$ and
$\leqslant 84\%$, respectively). This potential has been used in a large number
of studies of accretion flows onto non-rotating black holes to mimic essential
general relativistic effects within a Newtonian framework \citep[e.g.][]{
matsumoto,abramowicz88,chakrabarti95,macfadyen,hawley02,lee06,rosswog09}.

Other pseudo-Newtonian potentials have been introduced to give better
approximations to specific general relativistic features but at the price of
reproducing some other properties with less accuracy. For instance, the
pseudo-Newtonian potential introduced by \cite{nw} (NW hereafter)
\begin{equation}
\Phi_\s{\mathrm{NW}}(r) = -\frac{\G M}{r}
\left(1-\frac{3\,\rg}{r}+\frac{12\,r^2_\s{\mathrm{g}}}{r^2}\right),
\label{e1.2} 
\end{equation}
reproduces the angular frequencies $\Omega$ and $\Omega^\parallel$ with better
accuracy (p.e.~$\leqslant 14\%$ and $\leqslant 42\%$, respectively) than
$\Phi_\s{\mathrm{PW}}$ while still giving the correct location of
$r_\s{\mathrm{ms}}$. However, it locates $r_\s{\mathrm{mb}}$ at $\sim 3.5\,\rg$
and gives a less accurate estimate of the angular momentum of circular orbits
(p.e.~$\leqslant 30\%$). See Table~\ref{table1} for a general comparison of the
accuracy with which various relativistic properties are reproduced by
$\Phi_\s{\mathrm{N}}$, $\Phi_\s{\mathrm{PW}}$, and $\Phi_\s{\mathrm{NW}}$. Also
see \cite{artemova} for a comparison of the performance of
$\Phi_\s{\mathrm{PW}}$, $\Phi_\s{\mathrm{NW}}$, and other pseudo-Newtonian
potentials in reproducing the structure of a thin disc around rotating and
non-rotating black holes.
   
Pseudo-Newtonian potentials have been widely used in accretion studies, although
their range of applicability has been limited by the fact that no single one of
them can reproduce equally well all of the various dynamical properties of
Schwarzschild spacetime. For instance, a poor estimation of the orbital and
epicyclic frequencies hampers their ability to capture time-dependent behaviour,
such as the onset of instabilities in an accretion disc that might eventually
lead to observed signatures \citep[such as quasi-periodic oscillations; see
e.g.][]{kato01}. On the other hand, a large error in the estimation of the
binding energy of Keplerian circular orbits will lead to inaccurate estimates of
the total luminosity of accretion discs. Another issue that is commonly
overlooked is that most pseudo-Newtonian potentials are designed for accurately
reproducing circular orbits but not necessarily more general trajectories.
Nevertheless, they are frequently used in applications in which correctly
reproducing general trajectories might be of crucial importance (e.g.~the
collapsing interior of a massive star towards a newborn black hole or successive
passages of a star orbiting a black hole before becoming tidally disrupted).

In this work, we propose a generalization of the Newtonian potential that
accurately describes the motion of test particles in Schwarzschild spacetime
while still being formulated in entirely Newtonian
language.\footnote{Interestingly, \cite{abramowicz97} arrived at an equivalent
formulation from a different approach in which they considered Newtonian gravity
in a curved space. However, they did not explore general particle motion in
their work.} In addition to the Newtonian $1/r$-term, our potential includes an
explicit dependence on the velocity of the test particle. Following the
nomenclature of Lagrangian mechanics \citep[see e.g.][]{goldstein}, we therefore
call it a generalized Newtonian potential.\footnote{The potential introduced by
\cite{semerak_karas} for a Newtonian description of test particle motion around
a rotating black hole is another example of such a generalized potential.
However, when this potential is applied to the non-rotating case, it does not
give an overall satisfactory performance (e.g.~one finds $r_\s{\mathrm{mb}} =
r_\s{\mathrm{ms}} = 0$).} Moreover, this potential is derived from the actual
geodesic motion of test particles in Schwarzschild spacetime. This is in
contrast to most pseudo-Newtonian potentials that are introduced as ad hoc
recipes or as fitting formulae that mimic certain relativistic features
(see, nevertheless, \citealt{abramowicz09} for an a posteriori derivation of
$\Phi_\s{\mathrm{PW}}$). 

The remainder of the paper is organized as follows. In Section~\ref{s2}, the
generalized potential and the corresponding equations of motion are derived and
contrasted with the exact relativistic expressions. In Section~\ref{s3}, we
compare the performance of our generalized potential with $\Phi_\s{\mathrm{N}}$,
$\Phi_\s{\mathrm{PW}}$, and $\Phi_\s{\mathrm{NW}}$ in reproducing several
dynamical features of test particle motion in Schwarzschild spacetime including
purely radial infall, Keplerian and non-Keplerian circular motion, general
trajectories, pericentre advance, and two simple analytic accretion models (the
thin disc model of \citealt{shakura} as generalized to Schwarzschild spacetime
by \citealt{novikov73} and the toy accretion model of \citealt{tejeda2}). At the
end of this section we implement the new potential within a Newtonian smoothed
particle hydrodynamics (SPH) code and simulate the tidal disruption of a
solar-type star by a supermassive black hole. Finally, we summarize our results
in Section~\ref{s4}. 

\setcounter{equation}{0}
\section{Generalized Newtonian potential}
\label{s2} 

Consider a test particle with four-velocity\footnote{Greek indices run over
spacetime components, Latin indices run only over spatial components and the
Einstein summation convention over repeated indices is adopted.}  $u^\s{\mu} =
\ud x^\s{\mu}/\ud \tau$ following a timelike geodesic in Schwarzschild spacetime
(where $\tau$ is the proper time as measured by a comoving observer). Given the
staticity and spherical symmetry of the Schwarzschild metric, the motion of the
particle is restricted to a single plane (orbital plane) and is characterized by
the existence of two first integrals of motion: its specific energy
$\mathcal{E}$ and its specific angular momentum $h_\s{\mathrm{S}}$ given by
\citep[e.g.][]{novikov}
\begin{gather}
\mathcal{E} = -u_t =  \cc^2\left(1-\frac{2\, \rg}{r}\right)\,
\frac{\ud t}{\ud \tau},
\label{e2.1} \\
h_\s{\mathrm{S}} = u_\varphi = r^2\frac{\ud \varphi}{\ud \tau},
\label{e2.2} 
\end{gather}
where $\varphi$ is an angle measured within the orbital plane. These two
equations together with the normalization condition of the four-velocity,
$u_\s{\mu}u^\s{\mu} = - \cc^2$, lead to the equation governing the radial motion
\begin{equation}
\left(\frac{\ud r}{\ud \tau}\right)^2 = \frac{\mathcal{E}^2- \cc^4}{\cc^2}
+\frac{2\, \G M}{r}-\frac{h^2_\s{\mathrm{S}}}{r^2}
\left(1-\frac{2\, \rg   }{r} \right) .
\label{e2.3} 
\end{equation}
Using \eq{e2.1}, we can rewrite \eqs{e2.2} and \eqref{e2.3} in terms of
derivatives with respect to the coordinate time $t$, i.e.
\begin{gather}
\frac{\ud \varphi}{\ud t} =  \frac{\cc^2}{\mathcal{E}}\left(1 - \frac{2\,
\rg}{r}\right) \frac{h_\s{\mathrm{S}}}{r^2}, 
\label{e2.4} \\
\frac{\ud r}{\ud t} = 
\frac{\cc^2}{\mathcal{E}}\left(1 - \frac{2\, \rg}{r}\right) 
\sqrt{2\,E_\s{\mathrm{S}} +\frac{2\, \G M}{r}-\frac{h^2_\s{\mathrm{S}}}{r^2}
\left(1-\frac{2\,\rg}{r} \right) },
\label{e2.5} 
\end{gather}
where 
\begin{equation}
E_\s{\mathrm{S}} = \frac{\mathcal{E}^2- \cc^4}{2\,\cc^2}.
\label{es} 
\end{equation}
This is a suitable definition of energy since, in the non-relativistic limit
(nrl) in which $ r/\rg \gg 1$ and $v^2/\cc^2 \ll 1$ (where $v$ can be either
$\dot{r}$ or $r\dot{\varphi}$ and the dot denotes differentiation with respect
to the coordinate time $t$), it converges to the specific Newtonian mechanical
energy, i.e.~
\begin{equation}
E_\s{\mathrm{S}} \xrightarrow[{\rm nrl}]{} E_\s{\mathrm{N}} \equiv
\frac{1}{2}\left(\dot{r}^2 +r^2\dot{\varphi}^2\right) -\frac{ \G M}{r} .
\label{e2.6}
\end{equation}

Our starting point for a generalization of the Newtonian potential is the
low-energy limit (lel) of \eq{es} where $\mathcal{E}\simeq \cc^2$ or,
equivalently, $E_\s{\mathrm{S}}\simeq 0 $, i.e.
\begin{equation}
E_\s{\mathrm{S}} \xrightarrow[{\rm lel}]{} 
E_\s{\G} \equiv  
\frac{1}{2}\left[\frac{r^2 \dot{r}^2}{(r-2\, \rg)^2} +
 \frac{r^3 \dot{\varphi}^2}{r-2\,\rg} \right]-\frac{ \G M}{r}.
\label{e2.7} 
\end{equation}
Note that this limit does not necessarily imply low velocities or weak field.
\eq{e2.7} can be recast as
\begin{equation}
E_\s{\G} = T +\Phi_\s{\G}-
\dot{r}\,\frac{\partial\Phi_\s{\G}}{\partial \dot{r}}-
\dot{\varphi}\,\frac{\partial\Phi_\s{\G}}{\partial \dot{\varphi}},
\label{e2.8}
\end{equation}
where $T=\left(\dot{r}^2+r^2\dot\varphi^2\right)/2$ is the non-relativistic
kinetic energy per unit mass and $\Phi_\s{\G}$ is a generalized Newtonian
potential given by
\begin{equation}
\Phi_\s{\G}(r,\dot{r},\dot{\varphi}) = - \frac{ \G M}{r} -
\left(\frac{ 2\,\rg }{r-2\,\rg}\right)
\left[\left(\frac{r-\rg}{r-2\,\rg} \right)\dot{r}^2
+\frac{r^2\dot{\varphi}^2}{2}\right]  .
\label{e2.9} 
\end{equation}
The first term on the right-hand side of \eq{e2.9} is the usual Newtonian
potential that dictates the gravitational attraction due to the interaction of
the central mass and the rest mass of a test particle, while the second term
can be interpreted as an additional contribution due to the kinetic energy 
being also gravitationally attracted by the central mass. Contrary to a
pseudo-Newtonian potential, $\Phi_\s{\G}$ does satisfy the Poisson equation
(when the only source of the gravitational field is the central mass). 

We now use $\Phi_\s{\G}$ to construct the following Lagrangian (per unit mass):
\begin{equation}
L =T-\Phi_\s{\G}  =
\frac{1}{2}\left[\frac{r^2 \dot{r}^2}{(r-2\, \rg)^2} +
 \frac{r^3 \dot{\varphi}^2}{r-2\,\rg} \right]+ \frac{ \G M}{r}.
\label{e2.10} 
\end{equation}
Since $L$ is independent of $t$, the energy $E_\s{\G}$ as defined in \eq{e2.7}
is indeed a conserved quantity of the corresponding evolution equations. On the
other hand, the independence of $L$ from $\varphi$ guarantees that the specific
angular momentum defined as
\begin{equation}
h_\s{\G}=\frac{\partial L}{\partial \dot{\varphi}} =
\frac{r^3\dot{\varphi}}{r-2\, \rg}
\label{e2.11} 
\end{equation}
is also conserved. By combining \eqs{e2.7} and \eqref{e2.11} we get the
following expression for the radial motion:
\begin{equation}
\frac{\ud r}{\ud t} = \left(1-\frac{2\, \rg}{r} \right)
\sqrt{2\,E_\s{\G} +\frac{2\, \G M}{r}
-\frac{h^2_\s{\G}}{r^2}\left(1-\frac{2\,\rg}{r} \right)},
\label{e2.12} 
\end{equation}
which has a clear resemblance to the exact relativistic expression in \eq{e2.5}
and, as we shall show below, correctly reproduces a number of relativistic
features. In particular, and in full consistency with the low-energy limit, note
that for particles for which $\mathcal{E}= \cc^2$ (i.e.~parabolic-like
energies), \eqs{e2.11} and \eqref{e2.12} are identical to their relativistic
counterparts (Eqs.\,\ref{e2.4} and \ref{e2.5}, respectively).

Even though the whole evolution of the test particle motion is already
determined by \eqs{e2.11} and \eqref{e2.12}, it is also useful to compute the
corresponding expressions for the accelerations coming from the Euler-Lagrange
equations for an arbitrary coordinate system $(r,\,\theta,\,\phi)$,
i.e.\footnote{The angular velocities $\dot\theta$ and $\dot\phi$ are simply
related to $\dot\varphi$ by the relation $\dot\varphi^2 = \dot\theta^2 +
\sin^2\theta\,\dot\phi^2 $.} 
\begin{align}
\ddot{r}  = & -\frac{ \G M}{r^2}\left(1-\frac{2\, \rg}{r}\right)^2 
+ \frac{2\, \rg   \,\dot{r}^2}{r(r-2\,\rg)} \nonumber \\
 &+ \left(r-3\,\rg\right)
\left(\dot\theta^2 + \sin^2\theta\,\dot\phi^2\right) , 
\label{e2.16} \\
\ddot{\theta} = &
-\frac{2\,\dot{r}\,\dot{\theta}}{r}\left(\frac{r-3\, \rg }{r-2\, \rg   }
\right)
+\sin\theta\cos\theta\,\dot\phi^2,
\label{e2.17}  \\
\ddot{\phi} = &
-\frac{2\,\dot{r}\,\dot{\phi}}{r}\left(\frac{r-3\, \rg }{r-2\, \rg}\right)
-2\,\cot\theta\,\dot\phi\,\dot\theta.
\label{e2.18} 
\end{align}
These equations should be compared against the general relativistic ones which
are given by
\begin{align}
\ddot{r}  = & -\frac{ \G M}{r^2}\left(1-\frac{2\, \rg}{r}\right)^2 
\frac{\cc^4}{\mathcal{E}^2}
+ \frac{2\, \rg   \,\dot{r}^2}{r(r-2\,\rg)} \nonumber \\
 &+ \left(r-3\,\rg\right)
\left(\dot\theta^2 + \sin^2\theta\,\dot\phi^2\right) , 
\label{e2.19} \\
\ddot{\theta} = &
-\frac{2\,\dot{r}\,\dot{\theta}}{r}\left(\frac{r-3\, \rg }{r-2\, \rg}\right)
+\sin\theta\cos\theta\,\dot\phi^2,
\label{e2.20}  \\
\ddot{\phi} = &
-\frac{2\,\dot{r}\,\dot{\phi}}{r}\left(\frac{r-3\, \rg }{r-2\,\rg}\right)
-2\,\cot\theta\,\dot\phi\,\dot\theta,
\label{e2.21} 
\end{align}
from where we can see that they are identical except for the factor
$\cc^4/\mathcal{E}^2$ multiplying the first term in \eq{e2.19}.

For an implementation of the present approach within an existing hydrodynamics
code, the acceleration components may be needed in Cartesian coordinates. The
corresponding expressions are provided in Appendix \ref{AA}.

\section{Comparison with previous approaches}
\label{s3} 

In the following subsections we compare the performance of
$\Phi\s{\mathrm{PW}}$, $\Phi_\s{\mathrm{NW}}$, and $\Phi_\s{\G}$
(Eqs.\,\ref{e1.1}, \ref{e1.2} and \ref{e2.9}, respectively) in reproducing
several relativistic features of the motion of test particles in Schwarzschild
spacetime. As a reference to illustrate the importance of relativistic effects,
we also show the results obtained by applying the Newtonian potential
$\Phi_\s{\mathrm{N}}$.

\begin{figure}
\begin{center}
  \includegraphics[width=84mm]{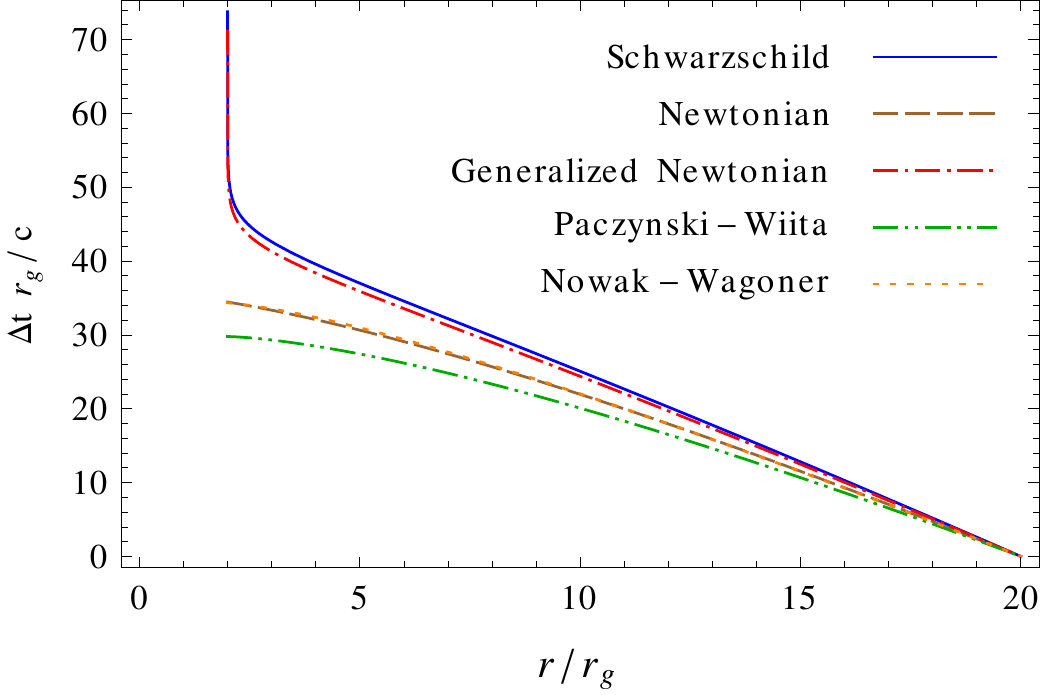}
\end{center}
 \caption{Infall time for a test particle falling radially from infinity as
calculated in general relativity and using the potentials $\Phi_\s{\mathrm{N}}$,
$\Phi_\s{\G}$, $\Phi_\s{\mathrm{PW}}$, and $\Phi_\s{\mathrm{NW}}$. In all the
cases we have taken $\dot{r}_\s{\infty} = -0.3\,\cc$ and have chosen $r=20\,\rg$
as the synchronization radius. Note that both $\Delta t_\s{\G}$ and $\Delta
t_\s{\mathrm{S}}$ diverge to infinity as the particle approaches the black hole
horizon located at $r=2\,\rg$.}
\label{f1}
\end{figure}

\begin{figure*}
\begin{center}
    \includegraphics[width=176mm]{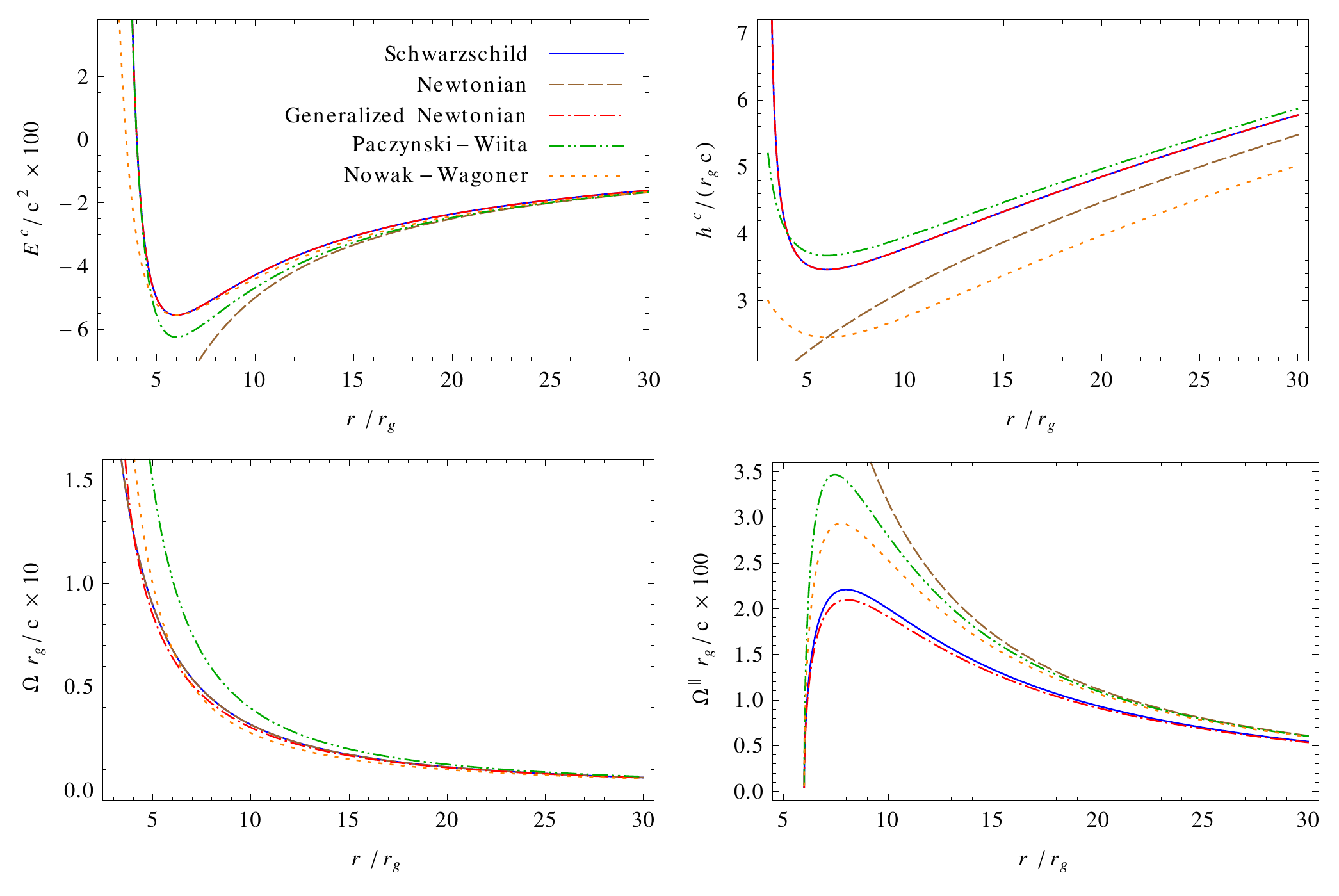} 
\end{center}
 \caption{Comparison of different quantities associated with circular motion.
From left to right and top to bottom the panels show the specific energy
$E^{c}$, the specific angular momentum $h^{c}$, the orbital angular velocity
$\Omega$, and the epicyclic frequency for small perturbations parallel to the
orbital plane $\Omega^\parallel$. Note that the curves for
$E^{c}_\s{\mathrm{G}}$ and $E^{c}_\s{\mathrm{S}}$\,, $h^{c}_\s{\mathrm{G}}$ and
$h^{c}_\s{\mathrm{S}}$\,, and $\Omega_\s{\mathrm{N}}$ and $\Omega_\s{\mathrm{S}}$
lie on top of each other. Analytic expressions for the different quantities
plotted in this figure are collected in Appendix~\ref{AB}.} 
\label{f2}
\end{figure*}

\setcounter{equation}{0}
\subsection{Radial infall}
\label{s3.1} 

Consider a particle in radial free-fall, i.e.~$\dot{\phi}=\dot{\theta}=0$. From
\eq{e2.12} it follows that the amount of time that it takes for the particle to 
fall from a radius $r_\s{2}$ to a smaller radius $r_\s{1}$ is given by
\begin{equation}
\Delta t_\s{\G} = \int_{r_\s{1}}^{r_\s{2}}\left[\left(1-\frac{2\,\rg}{r}\right)
\sqrt{2\,E_\s{\G}+\frac{2\,\G M}{r} }\right]^{-1}\ud r.
\label{e3.1} 
\end{equation}
It is clear that $\Delta t_\s{\G}$ will coincide with the corresponding
relativistic value $\Delta t_\s{\mathrm{S}}$ (as calculated from Eq.\,\ref{e2.5}) 
only when $E_\s{\mathrm{S}}=E_\s{\G}=0$, i.e.~for a vanishing radial velocity at 
infinity. In the general case, we will have $\Delta t_\s{\G}>\Delta 
t_\s{\mathrm{S}}$ if $E_\s{\G}<0$ and $\Delta t_\s{\G}<\Delta t_\s{\mathrm{S}}$ 
if $E_\s{\G}>0$. Also note that both $\Delta t_\s{\G}$ and $\Delta 
t_\s{\mathrm{S}}$ diverge as $r_\s{1}\rightarrow 2\,\rg$, which coincides with 
the description made by an observer situated at spatial infinity in Schwarzschild 
spacetime. In Fig.~\ref{f1}, we show the infall time as calculated from the 
relativistic solution and compare it with those coming from the use of 
$\Phi_\s{\mathrm{N}}$, $\Phi_\s{\G}$, $\Phi_\s{\mathrm{PW}}$, and 
$\Phi_\s{\mathrm{NW}}$.

\subsection{Circular orbits}
\label{s3.2} 

We consider now the special case of circular orbits as defined by the conditions
\begin{equation}
\dot{r} = 0\quad\text{and}\quad \ddot{r} = 0.
\label{e4.1} 
\end{equation}
After substituting these two conditions into \eqs{e2.12} and \eqref{e2.16} we
get a system of two equations that can be solved for the corresponding values of
$h_\s{\G}$ and $E_\s{\G}$ as
\begin{gather}
h^c_\s{\G} =  \frac{\sqrt{ \G M}\,r}{\sqrt{r-3\, \rg }},
\label{e4.2} \\ 
E^c_\s{\G} = -\frac{\G M}{2\,r}
\left(\frac{r-4\,\rg}{r-3\,\rg} \right),
\label{e4.3} 
\end{gather}
which are identical to the relativistic expressions and lead to the exact
locations for the corresponding photon orbit $r_\s{\mathrm{ph}}=3\, \rg $ (at
which both $h^c_\s{\G}$ and $E^c_\s{\G}$ diverge), the marginally bound orbit
$r_\s{\mathrm{mb}}=4\, \rg $ (for which $E^c_\s{\G} = 0$), and the marginally
stable orbit $r_\s{\mathrm{ms}}=6\, \rg $ (at which both $h^c_\s{\G}$ and
$E^c_\s{\G}$ reach their minima). See the top panels of Figure~\ref{f2} for a
comparison of $E^c$ and $h^c$ as calculated using the different potentials.

On the other hand, by combining \eqs{e2.11} and \eqref{e4.2} one finds the
orbital angular velocity 
\begin{equation}
\Omega_\s{\G} = \sqrt{\frac{\G M}{r-3\,\rg}}\left(\frac{r-2\,\rg}{r^2}\right),
\label{e4.4} 
\end{equation}
which should be compared against the exact expression in Schwarzschild spacetime
\begin{equation}
\Omega_\s{\mathrm{S}} = \sqrt{\frac{ \G M}{r^3}}.
\label{e4.5} 
\end{equation}
For $r\ge 6\,\rg$, $\Omega_\s{\G}$ reproduces the exact value with an accuracy
better than $5.7\%$. In the bottom-left panel of Figure~\ref{f2} we compare
these two frequencies as well as the ones corresponding to the potentials
$\Phi_\s{\mathrm{N}}$, $\Phi_\s{\mathrm{PW}}$, and $\Phi_\s{\mathrm{NW}}$. The
analytic expressions of all of the different quantities plotted in this figure
are summarized in Appendix~\ref{AB}.

\subsection{Perturbed circular orbits}

We now calculate the epicyclic frequencies associated with a small perturbation
of a stable circular orbit in the equatorial plane. The unperturbed trajectory
satisfies 
\begin{align}
x^\s{i}(t) & = (r,\pi/2,\Omega_\s{\G}\,t),\nonumber\\
\dot{x}^\s{i}(t)& = (0,0,\Omega_\s{\G}),
\end{align}
while for the perturbed trajectory we have 
\begin{align}
{x'}^\s{i}(t) & =
(r+\delta r,\pi/2+\delta \theta, \Omega_\s{\G}\,t+\delta \phi), \nonumber\\
\dot{x'}^\s{i}(t) & = (\delta\dot{ r} , \delta\dot{ \theta},
\Omega_\s{\G}+ \delta\dot{ \phi}).
\end{align}
Substituting these expressions into \eqs{e2.16}-\eqref{e2.18} we obtain the
following system of (linearized) differential equations for the perturbed
quantities
\begin{align}
\delta\ddot{ r} & = 
\left[\Omega^2_\s{\G}+\frac{2\, \G
M}{r^5}(r-2\, \rg   )(r-4\, \rg )\right]\delta r \nonumber\\
& \hspace{12pt}  +2\,\Omega_\s{\G}(r-3\, \rg )\delta\dot{\phi},
\label{e4.6} \\
\delta\ddot{\theta} & = -\Omega^2_\s{\G}\,\delta\theta
\label{e4.7}  \\
\delta\ddot{ \phi} &
=-\frac{2\,\Omega_\s{\G}}{r}\left(\frac{r-3\, \rg }{r-2\, \rg   }
\right)
\delta\dot{ r}.
\label{e4.8} 
\end{align}
From \eq{e4.7} we see that, to first order, the vertical perturbation decouples
from the other two directions and has the same angular frequency as the orbital 
motion, i.e.
\begin{equation}
\Omega^\perp_\s{\G} = \Omega_\s{\G},
\label{e4.9} 
\end{equation}
as is also the case for Schwarzschild spacetime. On the other hand, from
\eqs{e4.6} and \eqref{e4.8} we can see that the radial and azimuthal
perturbations are coupled. Following \cite{semerak}, we assume that the solution
to both equations is a harmonic oscillator with common angular frequency
$\Omega^\parallel_\s{\G}$, i.e.~$\delta r =\delta r_\s{0}\, {\rm
e}^{i\,\Omega^\parallel_\s{\G}\,t}$ and $\delta \phi=\delta \phi_\s{0}\, {\rm
e}^{i\, \Omega^\parallel_\s{\G}\,t}$, where $\delta r_\s{0}$ and $\delta
\phi_\s{0}$ are constant amplitudes. After substituting these solutions into
\eqs{e4.6} and \eqref{e4.8} we get as the only non-trivial solution
\begin{equation}
\Omega^\parallel_\s{\G} =
\sqrt{\frac{ \G M}{r^5}\left(\frac{r-6\, \rg }{r-3\, \rg }\right)}
(r-2\,\rg),
\end{equation}
which should be compared against the exact relativistic value
\begin{equation}
\Omega^\parallel_\s{\mathrm{S}} = 
\sqrt{\frac{ \G M}{r^3}\left(1-\frac{6\,\rg}{r}\right)}.
\end{equation}
Again, for $r\ge 6\,\rg$, $\Omega^\parallel_\s{\G}$ reproduces the exact value
with an accuracy better than $5.7\%$ (see the bottom-right panel of
Figure~\ref{f2}).

Note that, even though none of the three frequencies $\Omega_\s{\G}$,
$\Omega^\perp_\s{\G}$, and $\Omega^\parallel_\s{\G}$ coincides with the
corresponding relativistic expressions, they do keep the same ratios, i.e.~
\begin{equation}
\frac{\Omega^\perp_\s{\G}}{\Omega_\s{\G}} =
\frac{\Omega^\perp_\s{\mathrm{S}}}{\Omega_\s{\mathrm{S}}} = 1
\qquad \text{and} \qquad
\frac{\Omega^\parallel_\s{\G}}{\Omega_\s{\G}} =
\frac{\Omega^\parallel_\s{\mathrm{S}}}{\Omega_\s{\mathrm{S}}} = 
\sqrt{1-\frac{6\,\rg}{r}}.
\end{equation}
The pseudo-Newtonian potential introduced by \cite{kluzniak} also reproduces
exactly the ratio $\Omega^\parallel/\Omega$, although it does not reproduce
satisfactorily other important features (e.g.~one obtains $r_\s{\mathrm{ms}}
=3\,\rg$, $r_\s{\mathrm{mb}}=1.3\,\rg$, and a p.e.~of $\sim 42\%$ for the
binding energy).

\subsection{Non-Keplerian circular orbits}
\label{s3.3} 

Consider a test particle going round a circular orbit with uniform angular
velocity $\Omega=\ud\phi/\ud t$. This velocity does not necessarily coincide
with the Keplerian value and, if it does not, an external radial force must be
continuously applied in order to keep it on this circular trajectory. The actual
nature of this force (e.g.~pressure gradients, electromagnetic or propulsion
from a rocket) is unimportant for the present discussion. We can calculate the
necessary radial thrust from \eq{e2.16} as 
\begin{equation}
\mathcal{F}_\s{\mathrm{G}} =\frac{\G M}{r^2}\left(1-\frac{2\,\rg}{r}\right)^2-
 \Omega^2(r-3\,\rg).
\label{e5.1} 
\end{equation}
On the other hand, the corresponding relativistic expression is given by
\begin{equation}
\mathcal{F}_\s{\mathrm{S}} = a^\s{\hat{r}} = 
\left(1-\frac{2\,\rg}{r}\right)^{\nicefrac{-1}{2}} a^\s{r},
\label{e5.2} 
\end{equation}
where $a^\s{\hat{r}}$ is the physical value of the radial component of the
four-acceleration $a^\s{\mu}=u^\s{\nu}u^\s{\mu}_\s{;\nu}$ (i.e.~as projected
along the radial direction of a local tetrad). Writing the four-velocity of the
test particle as $u^\s{\mu}=u^\s{t}(1,0,0,\Omega)$, it is simple to check that
the only non-vanishing component of the four-acceleration is 
\begin{equation}
a^\s{r} = \left(1-\frac{2\,\rg}{r}\right)
\frac{\G M/r^2-r\,\Omega^2}{1-2\,\rg/r-r^2\Omega^2/\cc^2},
\label{e5.3} 
\end{equation}
from where we can rewrite \eq{e5.2} as
\begin{equation}
\mathcal{F}_\s{\mathrm{S}} =\left[\frac{\G M}{r^2}-
\frac{\Omega^2(r-3\,\rg)}{1-2\,\rg/r-r^2\Omega^2/c^2}\right]
\left(1-\frac{2\,\rg}{r}\right)^{\nicefrac{-1}{2}}.
\label{e5.4} 
\end{equation}

\begin{figure}
\begin{center}
  \includegraphics[width=84mm]{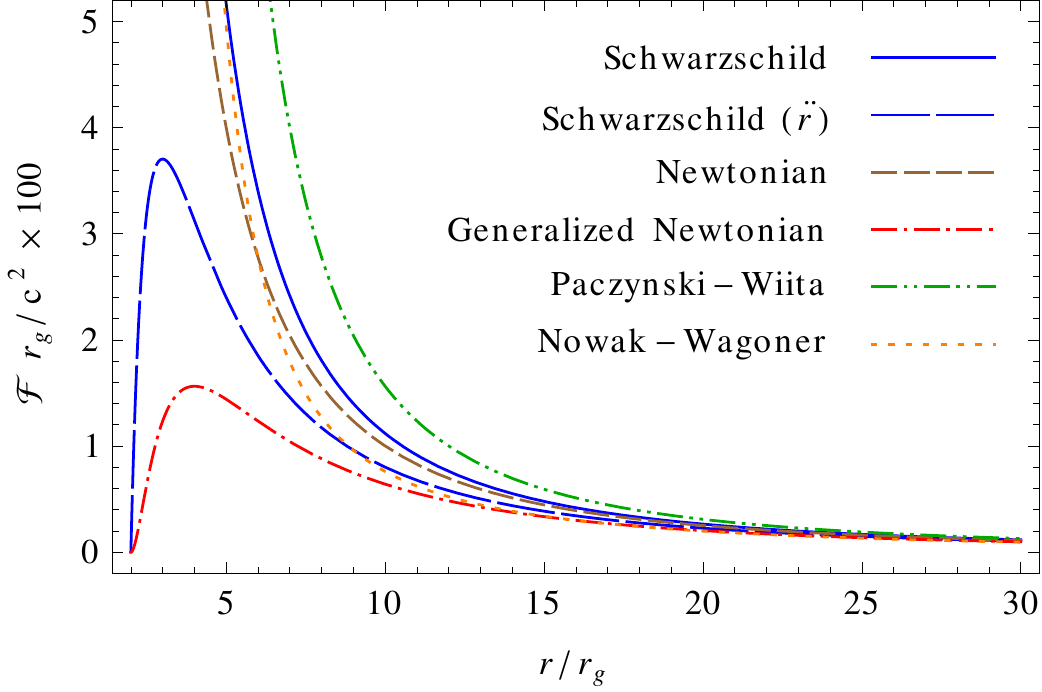}
\end{center}
 \caption{Radial thrust needed to keep a rocket hovering at a fixed radius. Note
that $\mathcal{F}_\s{\G}$ qualitatively follows the relativistic value of
$\ddot{r}$ and vanishes at the event horizon ($r=2\,\rg$). This is not a
physical result but rather a reflection of the fact that the Schwarzschild
coordinates are ill-behaved at this radius.}
\label{f3} 
\end{figure}

In Figure~\ref{f3} we compare $\mathcal{F}$ as obtained from the different
potentials for $\Omega=0$, i.e.~for a rocket that remains static at a fixed
radius. From this figure we see that $\mathcal{F}_\s{\G}$ gives the worst
approximation to the relativistic value (p.e.~$\leqslant 64\%$ for $r>6\,\rg$).
In particular, note that $\mathcal{F}_\s{\mathrm{S}}$ diverges to infinity as
$r\rightarrow2\,\rg$ while $\mathcal{F}_\s{\G}$ vanishes at this radius.
However, the behaviour of $\mathcal{F}_\s{\G}$ is in qualitative agreement with
the relativistic expression for $\ddot{r}\ (= \G M(r-2\,\rg)/r^3)$ which is a
clear indication that the description made using $\Phi_\s{\G}$ does not
correspond to a local observer but rather to one located at spatial infinity.

\begin{figure*}
\begin{center}
  \includegraphics[width=176mm]{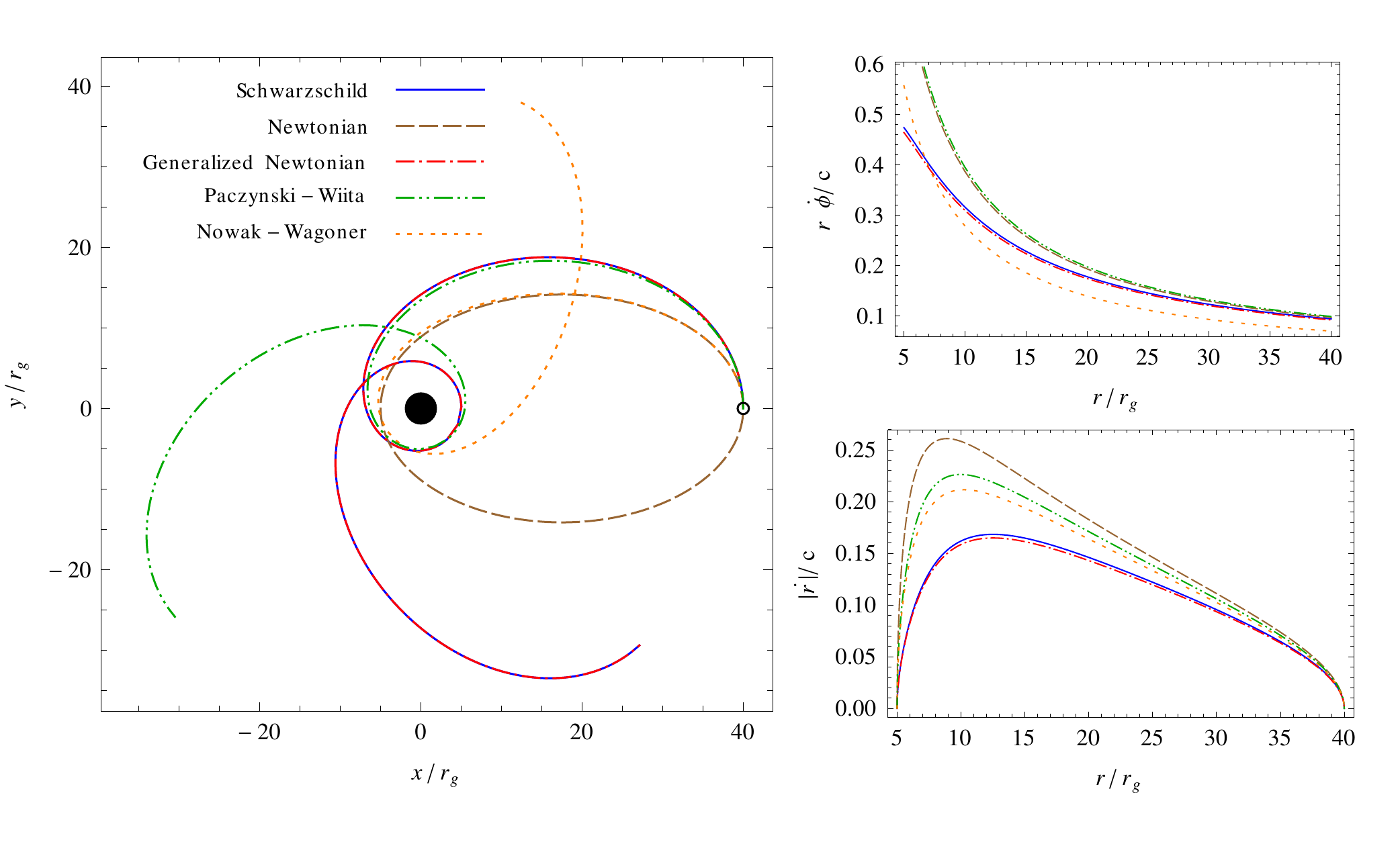}
\end{center}
 \caption{Comparison of an elliptic-like trajectory in Schwarzschild spacetime
with those obtained using  $\Phi_\s{\mathrm{N}}$, $\Phi_\s{\G}$,
$\Phi_\s{\mathrm{PW}}$, and $\Phi_\s{\mathrm{NW}}$. In all the cases, the
pericentre and apocentre of the orbit have been fixed as $r_\s{p} = 5\,\rg$ and
$r_\s{a} = 40\,\rg$, respectively. The common initial point is indicated by an
open circle. The left-hand panel shows the spatial projection of the different
trajectories onto the $x$-$y$ plane (with $x=r\cos\varphi$ and
$y=r\sin\varphi$). The panels on the right-hand side show the corresponding
azimuthal (top) and radial (bottom) velocities as functions of the radius. Note
that the spatial projection of the trajectory associated with the potential
$\Phi_\s{\G}$ reproduces exactly the relativistic one, while the velocities are
very well approximated.}
\label{f4}
\end{figure*}

Note that \eq{e5.1} captures an important feature of the exact relativistic
expression in \eq{e5.4}, namely that the radial thrust becomes independent of
the angular velocity at the location of the circular photon orbit $r=3\,\rg$
\citep{abramowicz_lasota}, and that the centrifugal force reverses sign at this
radius \mbox{\citep{abramowicz90}}. As \cite{abramowicz_miller} pointed out, a
consequence of this effect is the general relativistic result that the
ellipticity of a slowly rotating Maclaurin spheroid changes monotonicity at
$r\approx5\,\rg$ as the spheroid contracts with constant mass and angular
momentum (whereas in the Newtonian case the ellipticity monotonically increases
as the mean radius of the shrinking body decreases) \citep{chandra_miller,
miller77}. Following the same approach as in \cite{abramowicz_miller}, it is
simple to check that the present Newtonian description leads to the same
expression for the ellipticity as their Eq.\,(12') that quantitatively
reproduces this effect (p.e.~$\leqslant 35\%$ for $r>4\,\rg$).

\subsection{General orbits and pericentre advance}
\label{s3.4}

The geometric description of the trajectory followed by a test particle is
obtained by combining \eqs{e2.11} and \eqref{e2.12} as
\begin{equation}
\frac{\ud r}{\ud \varphi} = \frac{\sqrt{P(r)}}{h_\s{\G}} ,
\label{e6.1}
\end{equation}
where
\begin{equation}
P(r) = 2\,E_\s{\G}\,r^4 +2\,
\G M r^3-h^2_\s{\G}\,r\left(r-2\, \rg \right).
\label{e6.2}
\end{equation}
\eq{e6.1} is formally identical to the general relativistic expressions once the
correspondences
\begin{equation}
E_\s{\G}\leftrightarrow E_\s{\mathrm{S}} \qquad \text{and}
\qquad h_\s{\G}\leftrightarrow h_\s{\mathrm{S}}
\label{e2.13}
\end{equation} 
have been taken. This means that the spatial projection of the trajectories
obtained with $\Phi_\s{\G}$ is identical to that coming from the full
Schwarzschild solution. In other words, the orbital parameters (e.g.~apocentre,
pericentre, eccentricity) and characteristics (e.g.~pericentre advance, whether
it is bound, unbound or eventually trapped) have the same functional dependence
on the corresponding constants of motion. In Figure \ref{f4}, we show an example
of a generic elliptic-like trajectory as resulting from the full general
relativistic solution and compare it with those coming from the use of
$\Phi_\s{\mathrm{N}}$, $\Phi_\s{\G}$, $\Phi_\s{\mathrm{PW}}$, and
$\Phi_\s{\mathrm{NW}}$.

Just as in the general relativistic case, the solution to \eq{e6.1} can be
written in terms of elliptic integrals \citep[see e.g.][]{chandra, tejeda2}. In
the particular case of a bound orbit (elliptic-like trajectory) such that
$r\in[r_\s{p},r_\s{a}]$ (where $r_\s{p}$ and $r_\s{a}$ are the pericentre and
apocentre of the orbit, respectively), we define its (half) orbital period as
\begin{equation}
\Pi_\s{\G} = \int_{r_\s{p}}^{r_\s{a}} \frac{h_\s{\G}\,\ud r}{\sqrt{P(r)}}
 = \frac{2\,h_\s{\G}\,K(k)}{\sqrt{-2\,E_\s{\G}\,r_\s{p}(r_\s{a}-r_\s{b})}} ,
\label{e6.3} 
\end{equation}
where $K(k)$ is the complete elliptic integral of the first kind whose modulus
$k$ is given by
\begin{equation}
k = \sqrt{\frac{r_\s{b}(r_\s{a}-r_\s{p})}{r_\s{p}(r_\s{a}-r_\s{b})}} ,
\label{e6.4} 
\end{equation}
and $r_\s{b} = \G M/E_\s{\G}-r_\s{a}-r_\s{p}$ is the smallest positive root of
$P(r)$. The pericentre advance or precession is given by
\begin{equation}
\text{pericentre advance} = \Pi_\s{\G} - \pi.
\label{e6.5} 
\end{equation}
It is simple to check that, in the non-relativistic limit, $r_\s{b} \rightarrow
0$ and, thus, $k\rightarrow 0$. Given that $K(0)=\pi/2$, it follows then from
\eq{e6.3} that 
\begin{equation}
\Pi_\s{\G} \xrightarrow[{\rm nrl}]{} \pi.
\label{e6.6} 
\end{equation}

\begin{figure}
\begin{center}
  \includegraphics[width=84mm]{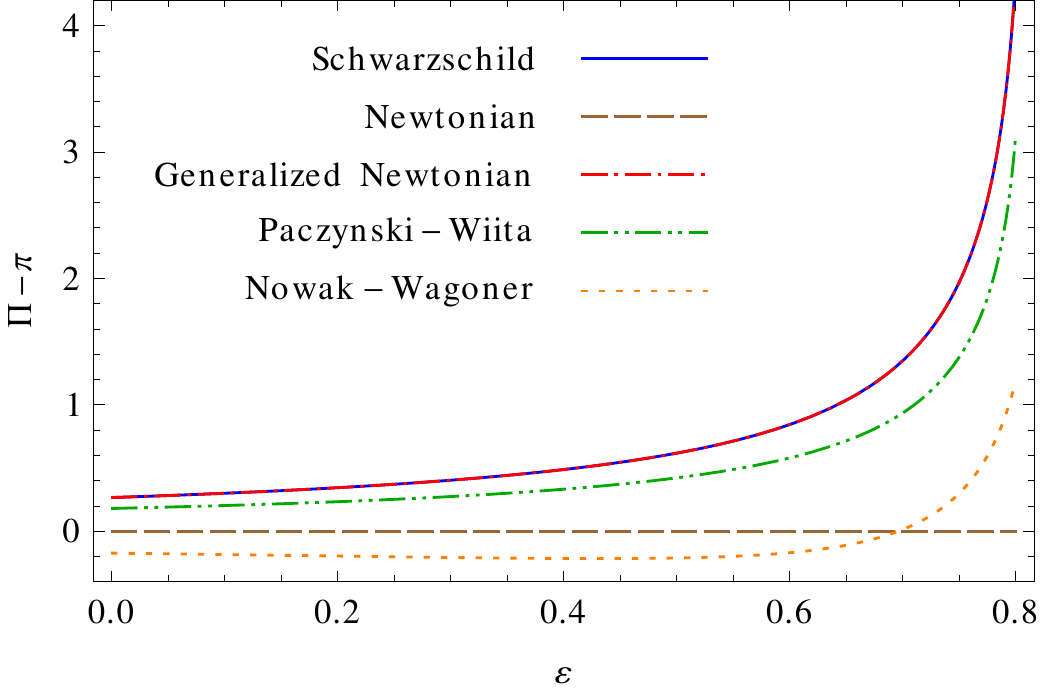}
\end{center}
 \caption{Pericentre advance as a function of the eccentricity $\varepsilon$ for
a fixed apocentre $r_\s{a}=40\,\rg$. Note that $\Phi_\s{\G}$ reproduces exactly
the relativistic value, while the result obtained with $\Phi_\s{\mathrm{PW}}$ is
consistently below the exact value by $\sim 30\%$. On the other hand, for most
values of $\varepsilon$, $\Phi_\s{\mathrm{NW}}$ produces a negative shift.}
\label{f5} 
\end{figure}

In Figure~\ref{f5}, we have plotted the pericentre advance as a function of the
eccentricity $\varepsilon=(r_\s{a}-r_\s{p})/(r_\s{a}+ r_\s{p})$ as predicted by
general relativity, $\Phi_\s{\mathrm{N}}$, $\Phi_\s{\G}$,
$\Phi_\s{\mathrm{PW}}$, and $\Phi_\s{\mathrm{NW}}$. As already mentioned, the
pericentre advance resulting from $\Phi_\s{\G}$ corresponds to the exact
relativistic value. On the other hand, the pericentre advance obtained from
$\Phi_\s{\mathrm{PW}}$ is off by $\sim 30\%$, while, for most values of the
eccentricity, $\Phi_\s{\mathrm{NW}}$ yields a negative shift (i.e.~a pericentre
lag rather than an advance).

The pseudo-Newtonian potential introduced by \cite{wegg} was specifically
designed to give an accurate description of the pericentre shift for
parabolic-like trajectories (p.e.~$\leqslant 1\%$). Nevertheless, this potential
is not well suited for studying more general cases (for instance, it does not
give the correct location of either $r_\s{\mathrm{ms}}$ or $r_\s{\mathrm{mb}}$).

\subsection{Accretion disc}
\label{s3.5}

\begin{figure}
\begin{center}
  \includegraphics[width=84mm]{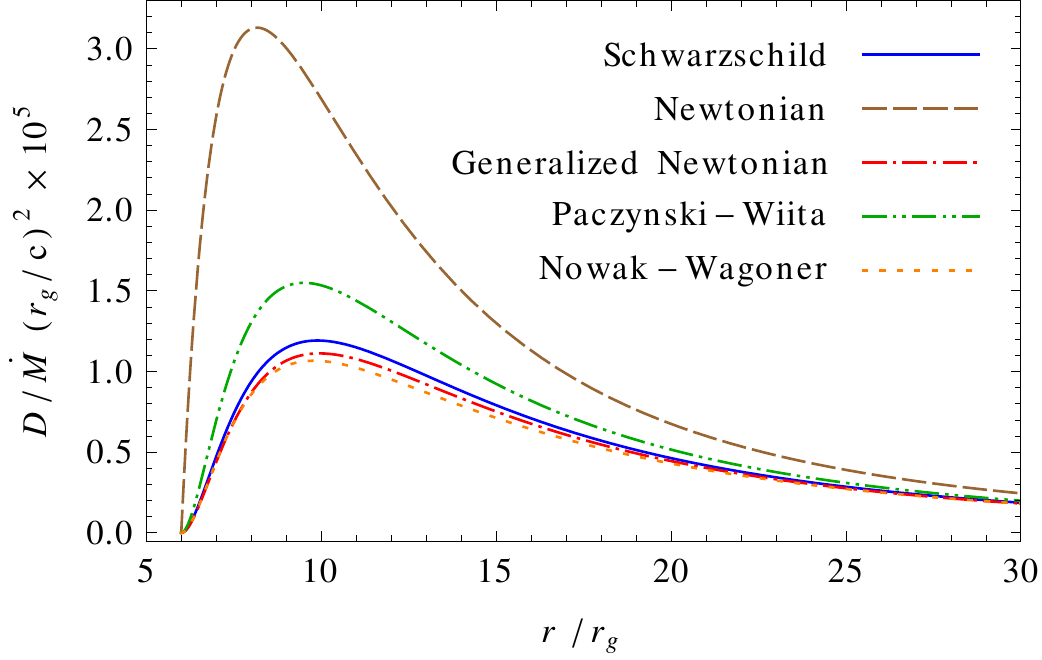}
\end{center}
 \caption{Energy flux per unit area $D$ as a function of radius emitted from the
surface of a geometrically thin, optically thick accretion disc.}
\label{f6} 
\end{figure}

In this section we consider the simple model of the stationary, geometrically
thin and optically thick accretion disc first investigated by \cite{shakura} in
a Newtonian context and its extension to a Schwarzschild spacetime by
\cite{novikov73}. Under the assumptions used there, it turns out that the energy
flux per unit area from the disc surface depends on the mass flux but not on the
details of the viscosity prescription and is given by \citep[see e.g.][]{king02}
\begin{equation}
D = \frac{\dot{M}}{4\,\pi\,r}\left|\frac{\ud \Omega}{\ud r}\right|
\left[h(r)-h(r_\s{\mathrm{in}})\right],
\label{e8.1} 
\end{equation}
where $\dot{M}$ is the constant accretion rate and $r_\s{\mathrm{in}}$ is the
inner boundary of the disc at which the viscous stresses are assumed to vanish.
In the standard thin disc model, one takes $r_\s{\mathrm{in}} =
r_\s{\mathrm{ms}} $. The total luminosity from the two faces of the disc is
obtained after integrating \eq{e8.1} over the whole disc, i.e.
\begin{equation}
L = 2\int_{r_\s{\mathrm{in}}}^\infty D(r)\, 2\,\pi\,r\,\ud r= e \dot{M}\cc^2 ,
\label{e8.2} 
\end{equation}
where $e$ is called the efficiency of the accretion process. In the following
list, we give the values of $e$ for the different potentials
\begin{equation}
\begin{split}
& e_\s{\mathrm{N}} = -E_\s{\mathrm{N}}(r_\s{\mathrm{in}})/\cc^2 = 
\frac{1}{12} \simeq 0.083, \\
& e_\s{\mathrm{PW}} = -E_\s{\mathrm{PW}}(r_\s{\mathrm{in}})/\cc^2 = 
\frac{1}{16} \simeq 0.062,\\
& e_\s{\mathrm{NW}} = -E_\s{\mathrm{NW}}(r_\s{\mathrm{in}})/\cc^2 = 
\frac{1}{18} \simeq 0.056,\\
& e_\s{\mathrm{G}} = -E_\s{\mathrm{G}}(r_\s{\mathrm{in}})/\cc^2 = 
\frac{1}{18} \simeq 0.056,\\
& e_\s{\mathrm{S}} = 1 - \mathcal{E}(r_\s{\mathrm{in}})/\cc^2 = 
1 - \frac{2\sqrt{2}}{3} \simeq 0.057,
\end{split}
\label{e8.3} 
\end{equation}
from where we see that $\Phi_\s{\mathrm{NW}}$ and $\Phi_\s{\G}$ give an equally
good approximation to the relativistic value. Nonetheless, for the actual radial
dependence of $D$, the latter provides a better approximation (p.e.~$\leqslant
7.2\%$) than the former (p.e.~$\leqslant 11\%$). In Figure~\ref{f6}, we compare
$D$ as obtained from general relativity with the results from the various
potentials.

\subsection{Accretion inflow}
\label{s3.6}

\begin{figure*}
\begin{center}
  \includegraphics[width=168mm]{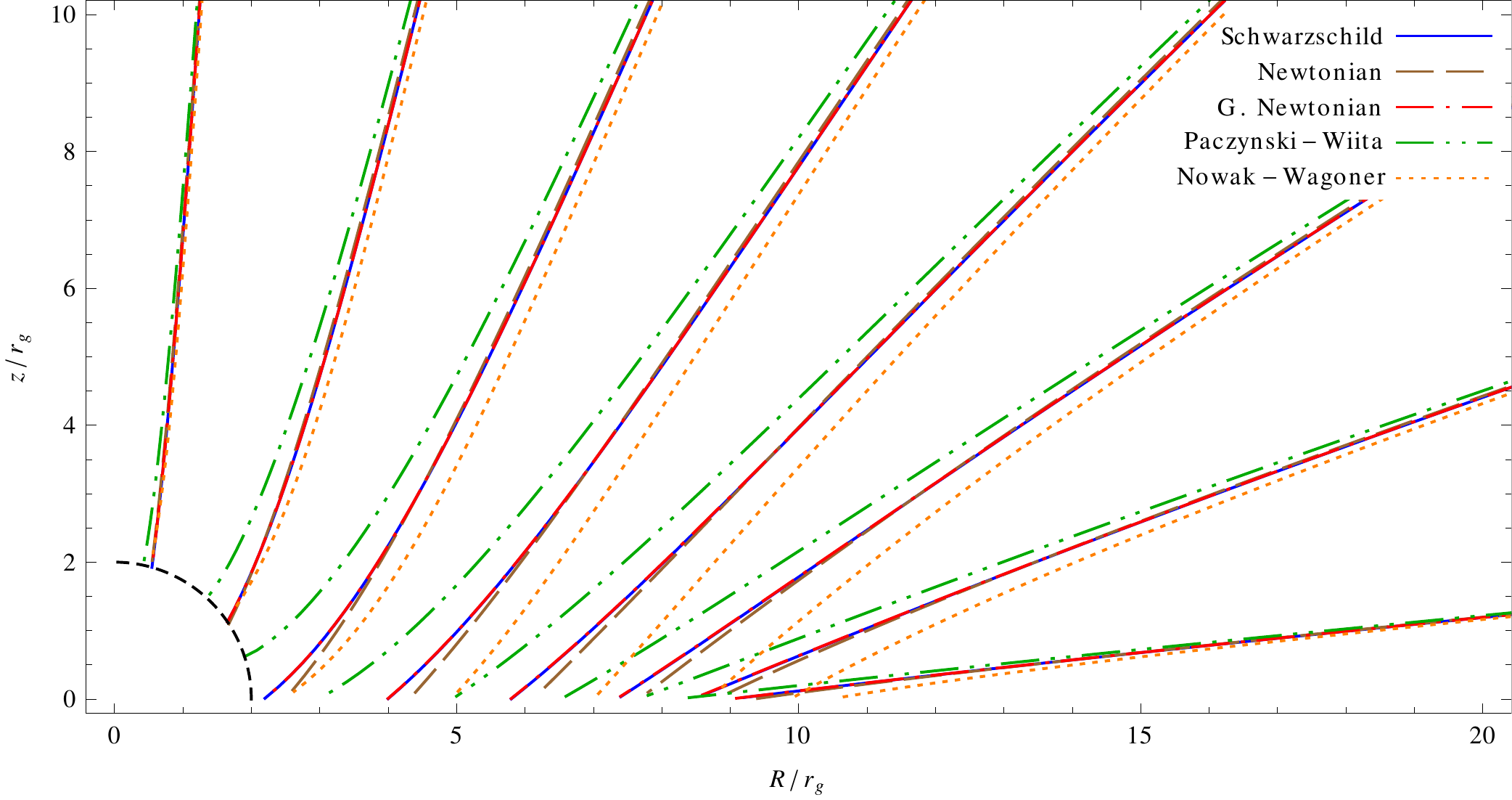}
\end{center}
 \caption{Streamlines of the steady-state accretion flow of a rotating gas cloud
of non-interacting particles onto a Schwarzschild black hole. The boundary
conditions are as given in \eq{e7.1}. The figure shows a zoom-in of the first 
quadrant of the $R$-$z$ plane (with $R=r\,\sin\theta$ and $z=r\,\cos\theta$). 
The black hole horizon is indicated with the dashed-line quarter circle.}
\label{f7} 
\end{figure*}

Here we consider the analytic accretion model of \cite{tejeda2} and
\cite{tejeda3}. In this model a rotating gas cloud of non-interacting particles
accretes steadily onto a Schwarzschild black hole. In Figure~\ref{f7}, we
compare the streamlines of the relativistic solution with the ones obtained from
the potentials $\Phi_\s{\mathrm{N}}$, $\Phi_\s{\G}$, $\Phi_\s{\mathrm{PW}}$, and
$\Phi_\s{\mathrm{NW}}$ for the following set of boundary conditions:
\begin{equation}
\begin{split}
M & = 4\,\Msun,\\
\dot{M} & = 0.01\, \Msun/s,\\
r_\s{0} & = 100\,\rg,\\
\dot{r}_\s{0} & = -0.141\,\cc,\\
r_\s{0}\,\dot{\phi}_\s{0} & = 0.038\,\cc,\\
\dot{\theta}_\s{0} & = 0. 
\end{split}
\label{e7.1}
\end{equation}
These boundary conditions are motivated by collapsing stellar cores leading to
long gamma-ray bursts and were used in \cite{tejeda2} to make a comparison with
one of the simulations of collapsar progenitors in \cite{lee06}.

Note that the streamlines obtained from $\Phi_\s{\G}$ are basically
indistinguishable from the exact relativistic ones. This is so because the
energies of the incoming fluid elements are close to parabolic,
i.e.~$\mathcal{E}\simeq \cc^2$ (for which case Eqs.\,\eqref{e2.11} and
\eqref{e2.12} coincide with the exact relativistic equations). With a different
choice of boundary conditions, the agreement with the relativistic results may
not be as good as in Figure~\ref{f7}, although $\Phi_\s{\G}$ is still in better
agreement with the relativistic solution than $\Phi_\s{\mathrm{PW}}$,
$\Phi_\s{\mathrm{NW}}$, or $\Phi_\s{\mathrm{N}}$.

\subsection{Tidal disruption}

\begin{figure*}
\begin{center}
  \includegraphics[width=168mm]{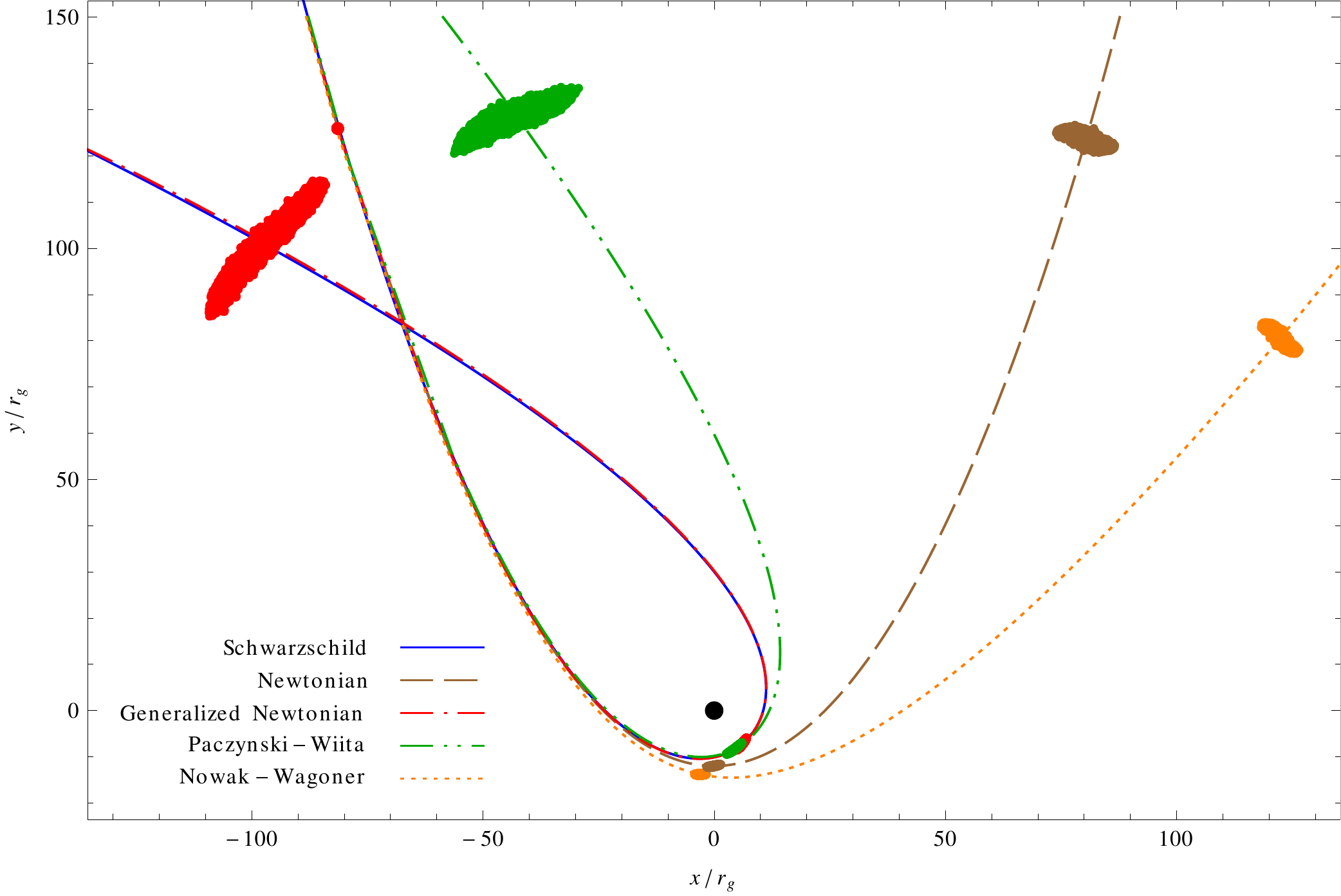}
\end{center}
 \caption{Tidal disruption of a solar-type star ($M_\ast= 1\,\Msun$, $R_\ast=
1\, \mathrm{R}_\odot$) by a supermassive black hole ($M_\s{\mathrm{BH}}= 10^6\,
\Msun$). The figure shows the outcome of four runs of the same SPH code for each
of the potentials $\Phi_\s{\mathrm{N}}$, $\Phi_\s{\G}$, $\Phi_\s{\mathrm{PW}}$,
and $\Phi_\s{\mathrm{NW}}$. A common initial configuration is used in which the
star is set at $r=200\,\rg$ along a parabolic trajectory with an encounter
strength of $\beta = 5$. Only $10^4$ SPH particles were used for each
simulation. The figure also shows the trajectory followed by the centre of mass
of the star in each case (as projected onto the orbital plane), together with
the projection of the SPH particles at three different points of the trajectory.
For reference, we also show with a solid blue line the exact geodesic trajectory
of a test particle in Schwarzschild spacetime.} 
\label{f8} 
\end{figure*}

In this section we apply the generalized potential to the tidal disruption of a
main-sequence star by a supermassive black hole. For doing this, we have
implemented the acceleration given in \eq{eab5} within the Newtonian SPH code
that we had used previously to simulate the tidal disruption of white dwarf
stars by moderately massive black holes \citep{rosswog09} and which has been
described in detail in \cite{rosswog08}. For recent reviews of the SPH method
see, e.g.~\cite{monaghan05} and \cite{rosswogsph}.

It is worth mentioning that for use in an Eulerian hydrodynamics code, it may be
beneficial to perform the simulation in the rest frame of the star to overcome
geometric restrictions and to minimize numerical artefacts due to advection
\citep[e.g.][]{cheng,guillochon13}. Such an approach may require a multipole
expansion of the tidal field around the stellar centre, but this is beyond the
scope of the current paper and left to future efforts.

Here we choose parameters that are identical to the ones of a general
relativistic simulation run presented in \cite{laguna93b}: the masses are
$M_\s{\mathrm{BH}}= 10^6\, \Msun$  for the black hole and $M_\ast= 1\,\Msun$ for
the star, the latter has a radius of $R_\ast= 1\, \mathrm{R}_\odot$ and is
modelled using a polytropic equation of state with exponent $\Gamma=5/3$. The
star approaches the central black hole following a parabolic trajectory with an
encounter strength $\beta= r_\s{\mathrm{t}}/r_\s{p}= 5$, where $r_\s{\mathrm{t}}
= (M_\s{\mathrm{BH}}/M_\ast)^{1/3}R_\ast \simeq 47\,\rg$ is the tidal radius and
$r_\s{p}\simeq 9.4\,\rg$ is the pericentre distance. As in the previous test
cases, we compare the results obtained with the generalized potential with those
using $\Phi_\s{\mathrm{N}}$, $\Phi_\s{\mathrm{PW}}$, and $\Phi_\s{\mathrm{NW}}$.

The results for the different approaches are displayed in Figure~\ref{f8}. For
ease of comparison with \cite{laguna93b}, we simply show the trajectory followed
by the centre of mass in each case together with the particle positions
projected onto the orbital plane at three different points of the trajectories.
Obviously, for such a deep encounter, relativistic effects lead to substantial
deviations from the Newtonian parabola. In this case, $\Phi_\s{\mathrm{PW}}$
produces a qualitatively correct result, although with only about $70\%$ of the
relativistic pericentre advance. As expected based on Fig.\,\ref{f5},
$\Phi_\s{\mathrm{NW}}$ gives a pericentre shift with the opposite sign to that
given by general relativity, and therefore leads to a wider orbit than the
purely Newtonian potential. Although the star passes the black hole with about
$0.4\,\cc$, the hydrodynamic result obtained from $\Phi_\s{\G}$ is in very close
agreement with the geodesic in Schwarzschild spacetime and the matter
distribution closely resembles the one shown in \cite{laguna93b} (see their
figure~1, second column).

\setcounter{equation}{0}
\section{Summary}
\label{s4}

We have derived a generalized Newtonian potential from the geodesic motion of a
test particle in Schwarzschild spacetime in the low-energy limit
$\mathcal{E}\simeq \cc^2$. In addition to the standard Newtonian term, the
generalized potential $\Phi_\s{\G}$ includes a term that depends on the square
of the velocity of the test particle and this can be interpreted as an
additional gravitational attraction from the central mass acting on the kinetic
energy of the test particle.

The new potential reproduces exactly several relativistic features of the motion
of test particles in Schwarzschild spacetime such as: the location of the
photon, marginally bound and marginally stable circular orbits; the radial
dependence of the energy and angular momentum of circular orbits; the ratio
between the orbital and epicyclic frequencies; the time evolution of
parabolic-like trajectories; the spatial projection of general trajectories as
function of the constants of motion and their pericentre advance. Moreover, the
equations of motion derived from this potential reproduce the reversal of the
centrifugal force at the location of the circular photon orbit \citep[see
e.g.][] {abramowicz90}. We are not aware of this relativistic feature ever
having been reproduced before by any pseudo-Newtonian potential. Additionally,
the equations obtained also provide a good approximation for the time evolution
of particles in free-fall, for the orbital angular velocity of circular orbits,
and for the epicyclic frequencies associated with small perturbations away from
circular motion (all of these corresponding to the description made by observers
situated far away from the central black hole). We have also applied
$\Phi_\s{\G}$ to the study of simple accretion scenarios, first for the thin
disc model of \cite{novikov73} and then for the accretion infall model of
\cite{tejeda2}, finding good agreement with the exact relativistic solutions in
both cases. 

As a further astrophysical application and a demonstration of the minimal effort
required to implement the suggested generalized potential within an existing
Newtonian hydrodynamics code, we have applied it to the tidal disruption of a
main-sequence star by a supermassive black hole. For this, we implemented the
equations of motion derived from $\Phi_\s{\G}$ within the 3D SPH code described
in \cite{rosswog08}. The results obtained are in very good agreement with the
relativistic simulation presented in \cite{laguna93b}.

\begin{table}
\centering
\begin{tabular}{ccccc}
\hline\hline\\[-7pt]
Potential & $\Phi_\s{\mathrm{N}}$ & $\Phi_\s{\mathrm{PW}}$ 
& $\Phi_\s{\mathrm{NW}}$ &  $\Phi_\s{\G}$
\\[1pt]\hline\\[-7pt]
$\Delta t(\dot{r}_\s{\infty} = 0)$ & $\leqslant 16.7\%$ 
& $\leqslant 23.8\%$ & $\leqslant 9.3\%$ &  
exact \\[2pt]
$\Delta t(\dot{r}_\s{\infty} = -0.3\,c)$ & $\leqslant 15.9\%$ 
& $\leqslant 24.3\%$ & $\leqslant 15\%$ &  
$\leqslant 2.9\%$ \\[2pt]
$r_\s{\mathrm{ph}}$ & $-$ & $33.3\%$ & $-$ &  exact \\[2pt]
$r_\s{\mathrm{mb}}$ & $-$ &  exact &  $13.4\%$ &  exact \\[2pt]
$r_\s{\mathrm{ms}}$ & $-$ &  exact &  exact &  exact \\[2pt]
$E^c$ & $\leqslant 50\%$ & $\leqslant 12.5\%$ & $\leqslant 3.3\%$ 
& exact \\[2pt]
$h^c$ & $\leqslant 29.3\%$ & $\leqslant 6.1\%$ & $\leqslant 29.3\%$ 
& exact \\[2pt]
$\Omega$ & exact & $\leqslant 50\%$ & $\leqslant 13.4\%$ 
& $\leqslant 5.7\%$ \\[2pt] 
$\Omega^\parallel$ & $\infty$ & $\leqslant 83.7\%$ & $\leqslant 41.4\%$ & 
$\leqslant 5.7\%$ \\[2pt]
$\mathcal{F}$ & $\leqslant 18.4\%$ & $\leqslant 83.7\%$ & $\leqslant 32.1\%$ 
& $\leqslant 63.7\%$ \\[2pt]
$\Pi - \pi $  & $-$ & $\sim 30\%$ & $>80\%$  & exact \\[2pt]
$e$ & $45.7\%$ & $9.3\%$ & $2.9\%$ & $2.9\%$ \\[2pt]
$D$ & $\infty$ & $\leqslant 58.9\%$ & $\leqslant 11.1\%$ 
& $\leqslant 7.3\%$ \\
\hline
\end{tabular}
\caption{Maximum percentage error obtained for the approximation of several
quantities associated with the motion of test particles in Schwarzschild
spacetime using the four potentials $\Phi_\s{\mathrm{N}}$,
$\Phi_\s{\mathrm{PW}}$, $\Phi_\s{\mathrm{NW}}$ and $\Phi_\s{\G}$. In all cases,
we have considered only $r>6\,\rg$. } 
\label{table1} 
\end{table}

In Table~\ref{table1}, we summarize the accuracy with which various relativistic
properties are reproduced by $\Phi_\s{\mathrm{N}}$, $\Phi_\s{\mathrm{PW}}$,
$\Phi_\s{\mathrm{NW}}$, and $\Phi_\s{\G}$. With the exception of the radial
thrust $\mathcal{F}$ needed to keep a rocket hovering at a static position, we
found that $\Phi_\s{\G}$ provides a more accurate description of the motion of
test particles in Schwarzschild spacetime than any of the other potentials. For
this reason, and given that a proper modelling of a realistic accretion scenario
requires that (at least) all of these relativistic features are accurately
reproduced, we consider the new potential as being a promising simple but 
powerful tool for studying many processes occurring in the vicinity of a 
Schwarzschild black hole.

\section{Acknowledgements}

It is a pleasure to thank John C.~Miller for insightful discussion and critical 
comments on the manuscript. We also thank Marek Abramowicz, Oleg Korobkin,
Iv\'an Zalamea and the anonymous referee for useful comments and suggestions.
The simulations of this paper were in part performed on the facilities of the 
H\"ochstleistungsrechenzentrum Nord (HLRN). This work has been supported by the 
Swedish Research Council (VR) under grant 621-2012-4870.

\bibliographystyle{mn2e}
\bibliography{references}

\begin{thebibliography}{}

\bibitem[\protect\citeauthoryear{{Abramowicz}}{{Abramowicz}}{1990}]{abramowicz%
90}
{Abramowicz} M.~A.,  1990, Monthly Notices of the Royal Astronomical Society,
  245, 733

\bibitem[\protect\citeauthoryear{{Abramowicz}}{{Abramowicz}}{2009}]{abramowicz%
09}
{Abramowicz} M.~A.,  2009, \aap, 500, 213

\bibitem[\protect\citeauthoryear{{Abramowicz}, {Czerny}, {Lasota} \&
  {Szuszkiewicz}}{{Abramowicz} et~al.}{1988}]{abramowicz88}
{Abramowicz} M.~A.,  {Czerny} B.,  {Lasota} J.~P.,    {Szuszkiewicz} E.,  1988,
  Astrophysical Journal, 332, 646

\bibitem[\protect\citeauthoryear{{Abramowicz}, {Lanza}, {Miller} \&
  {Sonego}}{{Abramowicz} et~al.}{1997}]{abramowicz97}
{Abramowicz} M.~A.,  {Lanza} A.,  {Miller} J.~C.,    {Sonego} S.,  1997,
  General Relativity and Gravitation, 29, 1583

\bibitem[\protect\citeauthoryear{{Abramowicz} \& {Lasota}}{{Abramowicz} \&
  {Lasota}}{1986}]{abramowicz_lasota}
{Abramowicz} M.~A.,  {Lasota} J.~P.,  1986, American Journal of Physics, 54,
  936

\bibitem[\protect\citeauthoryear{{Abramowicz} \& {Miller}}{{Abramowicz} \&
  {Miller}}{1990}]{abramowicz_miller}
{Abramowicz} M.~A.,  {Miller} J.~C.,  1990, Monthly Notices of the Royal
  Astronomical Society, 245, 729

\bibitem[\protect\citeauthoryear{{Artemova}, {Bj\"{o}rnsson} \&
  {Novikov}}{{Artemova} et~al.}{1996}]{artemova}
{Artemova} I.~V.,  {Bj\"{o}rnsson} G.,    {Novikov} I.~D.,  1996, Astrophysical
  Journal, 461, 565

\bibitem[\protect\citeauthoryear{{Chakrabarti} \& {Titarchuk}}{{Chakrabarti} \&
  {Titarchuk}}{1995}]{chakrabarti95}
{Chakrabarti} S.,  {Titarchuk} L.~G.,  1995, Astrophysical Journal, 455, 623

\bibitem[\protect\citeauthoryear{Chandrasekhar}{Chandrasekhar}{1983}]{chandra}
Chandrasekhar S.,  1983, The Mathematical Theory of Black Holes.
Oxford University Press

\bibitem[\protect\citeauthoryear{{Chandrasekhar} \& {Miller}}{{Chandrasekhar}
  \& {Miller}}{1974}]{chandra_miller}
{Chandrasekhar} S.,  {Miller} J.~C.,  1974, Monthly Notices of the Royal
  Astronomical Society, 167, 63

\bibitem[\protect\citeauthoryear{{Cheng} \& {Evans}}{{Cheng} \&
  {Evans}}{2013}]{cheng}
{Cheng} R.~M.,  {Evans} C.~R.,  2013, Physical Review D, 87, 104010

\bibitem[\protect\citeauthoryear{{Frank}, {King} \& {Raine}}{{Frank}
  et~al.}{2002}]{king02}
{Frank} J.,  {King} A.,    {Raine} D.,  2002, Accretion Power in Astrophysics,
  3rd edn.
Cambridge University Press

\bibitem[\protect\citeauthoryear{Frolov \& Novikov}{Frolov \&
  Novikov}{1998}]{novikov}
Frolov V.~P.,  Novikov I.~D.,  1998, Black Hole Physics: Basic Concepts and New
  Developments.
Kluwer Academic

\bibitem[\protect\citeauthoryear{{Goldstein}, {Poole} \& {Safko}}{{Goldstein}
  et~al.}{2002}]{goldstein}
{Goldstein} H.,  {Poole} C.,    {Safko} J.,  2002, {Classical Mechanics}, 3rd
  edn.
Addison-Wesley, San Francisco

\bibitem[\protect\citeauthoryear{{Guillochon} \& {Ramirez-Ruiz}}{{Guillochon}
  \& {Ramirez-Ruiz}}{2013}]{guillochon13}
{Guillochon} J.,  {Ramirez-Ruiz} E.,  2013, Astrophysical Journal, 767, 25

\bibitem[\protect\citeauthoryear{{Hawley} \& {Balbus}}{{Hawley} \&
  {Balbus}}{2002}]{hawley02}
{Hawley} J.~F.,  {Balbus} S.~A.,  2002, Astrophysical Journal, 573, 738

\bibitem[\protect\citeauthoryear{{Kato}}{{Kato}}{2001}]{kato01}
{Kato} S.,  2001, Publications of the Astronomical Society of Japan, 53, 1

\bibitem[\protect\citeauthoryear{{Klu{\'z}niak} \& {Lee}}{{Klu{\'z}niak} \&
  {Lee}}{2002}]{kluzniak}
{Klu{\'z}niak} W.,  {Lee} W.~H.,  2002, Monthly Notices of the Royal
  Astronomical Society, 335, L29

\bibitem[\protect\citeauthoryear{{Laguna}, {Miller}, {Zurek} \&
  {Davies}}{{Laguna} et~al.}{1993}]{laguna93b}
{Laguna} P.,  {Miller} W.~A.,  {Zurek} W.~H.,    {Davies} M.~B.,  1993,
  Astrophysical Journal Letters, 410, L83

\bibitem[\protect\citeauthoryear{{Lee} \& {Ram\'irez-Ruiz}}{{Lee} \&
  {Ram\'irez-Ruiz}}{2006}]{lee06}
{Lee} W.~H.,  {Ram\'irez-Ruiz} E.,  2006, Astrophysical Journal, 641, 961

\bibitem[\protect\citeauthoryear{{MacFadyen} \& {Woosley}}{{MacFadyen} \&
  {Woosley}}{1999}]{macfadyen}
{MacFadyen} A.~I.,  {Woosley} S.~E.,  1999, Astrophysical Journal, 524, 262

\bibitem[\protect\citeauthoryear{{Matsumoto}, {Kato}, {Fukue} \&
  {Okazaki}}{{Matsumoto} et~al.}{1984}]{matsumoto}
{Matsumoto} R.,  {Kato} S.,  {Fukue} J.,    {Okazaki} A.~T.,  1984,
  Publications of the Astronomical Society of Japan, 36, 71

\bibitem[\protect\citeauthoryear{{Miller}}{{Miller}}{1977}]{miller77}
{Miller} J.~C.,  1977, Monthly Notices of the Royal Astronomical Society, 179,
  483

\bibitem[\protect\citeauthoryear{{Monaghan}}{{Monaghan}}{2005}]{monaghan05}
{Monaghan} J.~J.,  2005, Reports on Progress in Physics, 68, 1703

\bibitem[\protect\citeauthoryear{{Novikov} \& {Thorne}}{{Novikov} \&
  {Thorne}}{1973}]{novikov73}
{Novikov} I.~D.,  {Thorne} K.~S.,  1973, in {C.~Dewitt \& B.~S.~Dewitt} ed.,
  Black Holes (Les Astres Occlus) {Astrophysics of black holes}.
Gordon and Breach, New York, pp 343--450

\bibitem[\protect\citeauthoryear{{Nowak} \& {Wagoner}}{{Nowak} \&
  {Wagoner}}{1991}]{nw}
{Nowak} M.~A.,  {Wagoner} R.~V.,  1991, Astrophysical Journal, 378, 656

\bibitem[\protect\citeauthoryear{{Paczy{\'n}sky} \& {Wiita}}{{Paczy{\'n}sky} \&
  {Wiita}}{1980}]{pw}
{Paczy{\'n}sky} B.,  {Wiita} P.~J.,  1980, Astronomy and Astrophysics, 88, 23

\bibitem[\protect\citeauthoryear{{Rosswog}}{{Rosswog}}{2009}]{rosswogsph}
{Rosswog} S.,  2009, New Astronomy Reviews, 53, 78

\bibitem[\protect\citeauthoryear{{Rosswog}, {Ram\'irez-Ruiz} \&
  {Hix}}{{Rosswog} et~al.}{2008}]{rosswog08}
{Rosswog} S.,  {Ram\'irez-Ruiz} E.,    {Hix} W.~R.,  2008, Astrophysical
  Journal, 679, 1385

\bibitem[\protect\citeauthoryear{{Rosswog}, {Ram\'irez-Ruiz} \&
  {Hix}}{{Rosswog} et~al.}{2009}]{rosswog09}
{Rosswog} S.,  {Ram\'irez-Ruiz} E.,    {Hix} W.~R.,  2009, Astrophysical
  Journal, 695, 404

\bibitem[\protect\citeauthoryear{{Semer{\'a}k} \& {Karas}}{{Semer{\'a}k} \&
  {Karas}}{1999}]{semerak_karas}
{Semer{\'a}k} O.,  {Karas} V.,  1999, \aap, 343, 325

\bibitem[\protect\citeauthoryear{{Semer{\'a}k} \& {{\v Z}{\'a}{\v
  c}ek}}{{Semer{\'a}k} \& {{\v Z}{\'a}{\v c}ek}}{2000}]{semerak}
{Semer{\'a}k} O.,  {{\v Z}{\'a}{\v c}ek} M.,  2000, Publications of the
  Astronomical Society of Japan, 52, 1067

\bibitem[\protect\citeauthoryear{{Shakura} \& {Sunyaev}}{{Shakura} \&
  {Sunyaev}}{1973}]{shakura}
{Shakura} N.~I.,  {Sunyaev} R.~A.,  1973, Astronomy and Astrophysics, 24, 337

\bibitem[\protect\citeauthoryear{{Tejeda}, {Mendoza} \& {Miller}}{{Tejeda}
  et~al.}{2012}]{tejeda2}
{Tejeda} E.,  {Mendoza} S.,    {Miller} J.~C.,  2012, Monthly Notices of the
  Royal Astronomical Society, 419, 1431

\bibitem[\protect\citeauthoryear{{Tejeda}, {Taylor} \& {Miller}}{{Tejeda}
  et~al.}{2013}]{tejeda3}
{Tejeda} E.,  {Taylor} P.~A.,    {Miller} J.~C.,  2013, Monthly Notices of the
  Royal Astronomical Society, 429, 925

\bibitem[\protect\citeauthoryear{{Wegg}}{{Wegg}}{2012}]{wegg}
{Wegg} C.,  2012, Astrophysical Journal, 749, 183

\end{thebibliography}

\appendix
\onecolumn

\setcounter{section}{0}
\setcounter{equation}{0}

\renewcommand{\theequation}{A.\arabic{equation}}

\section{Acceleration in Cartesian coordinates}
\label{AA}

The Cartesian coordinates $(x,y,z)$ are connected to the spherical ones
$(r,\theta,\phi)$ in the usual way, i.e.
\begin{align}
& x = r\,\sin\theta\cos\phi, & r & =\sqrt{x^2+y^2+z^2} ,\nonumber 
\\ & y = r\,\sin\theta\sin\phi, & 
\theta & = \tan^{-1}\left(\sqrt{x^2+y^2}/z\right), 
\label{eab1} 
\\ & z = r\,\cos\theta, & \phi & = \tan^{-1}\left(y/x\right),
\nonumber
\end{align}
from which we get
\begin{align}
r\,\dot{r} =\ & x\,\dot{x}+y\,\dot{y}+z\,\dot{z} 
= \sum_\s{i}\,x^\s{i}\,\dot{x}^\s{i},
\label{eab2} \\
r^4(\dot{\theta}^2+\sin^2\theta\,\dot{\phi}^2) =\ & 
(x\,\dot{y}-y\,\dot{x})^2 + (x\,\dot{z}-z\,\dot{x})^2 
+ (z\,\dot{y}-y\,\dot{z})^2 = \sum_i
 \left(\sum_\s{jk}\epsilon_\s{ijk}\,x^\s{j}\,\dot{x}^\s{k}\right)^2,
\label{eab3}
\end{align}
where $x^\s{i}=x,y,z$ and $\epsilon_\s{ijk}$ is the Levi-Civita symbol.
Substituting \eqs{eab2} and \eqref{eab3} into the Lagrangian in 
\eq{e2.10} gives
\begin{equation}
L = \frac{1}{2}\left[\left(\frac{\sum_\s{i} x^\s{i}\,\dot{x}^\s{i}}
{r-2\,\rg}\right)^2 + \frac{\sum_\s{i}\left(\sum_\s{jk}\epsilon_\s{ijk}\,
x^\s{j}\,\dot{x}^\s{k} \right)^2}{r(r-2\,\rg)}\right]  + \frac{ \G M}{r},
\label{eab4} 
\end{equation}
from which we get the following expression for the acceleration components
\begin{equation}
\ddot{x}^\s{i} =  -\frac{\G Mx^\s{i}}{r^3}\left(1-\frac{2\,\rg}{r}\right)^2 +
 \frac{2\,\rg\,\dot{x}^\s{i}}{r^2(r-2\,\rg)}\,\sum_j\,x^\s{j}\,\dot{x}^\s{j} 
 - \frac{3\,\rg\,x^\s{i}}{r^5}\,\sum_j
 \bigg(\sum_{kl}\epsilon_\s{jkl}\,x^\s{k}\,\dot{x}^\s{l}\bigg)^2.
 \label{eab5} 
\end{equation}

\section{Miscellaneous expressions}
\label{AB}

In the following table we collect the different formulae associated with
circular motion and plotted in Figures~\ref{f2} and \ref{f3}. 
The corresponding Newtonian expressions are also included to facilitate 
further comparison.

\begin{center}
\begin{tabular}{cccccc}
& Newton & Schwarzschild & Generalized Newtonian & Paczy\'{n}ski-Wiita &
Nowak-Wagoner
\\[2pt] \hline \\[-6pt]
$\mathcal{F}$ & $-\frac{ \G M}{r^2}$ & $-\frac{\G M}{\sqrt{r^3(r-2\,\rg)}}$ &
$-\frac{ \G M(r-2\,\rg)^2}{r^4}$ & $-\frac{ \G M}{(r-2\, \rg)^2}$ &
$-\frac{ \G M(r^2-6\,\rg \,r+36\,r^2_\s{g})}{r^4}$ \\[4pt]
 $(h^c)^2$ & ${\scriptstyle \G Mr}$ & 
$\frac{ \G Mr^2}{r-3\, \rg }$  &
$\frac{ \G Mr^2}{r-3\, \rg }$ &
$\frac{ \G Mr^3}{(r-2\, \rg   )^2}$ & 
$\frac{ \G M(r^2-6\, \rg \,r+36\,r^2_\s{g})}{r}$ \\[4pt]
 $E^c$ & $-\frac{ \G M}{2\,r}$ &
$-\frac{ \G M(r-4\, \rg )}{2\,r(r-3\, \rg )}$  &
$-\frac{ \G M(r-4\, \rg )}{2\,r(r-3\, \rg )}$ &
$-\frac{ \G M(r-4\, \rg )}{2(r-2\, \rg   )^2}$ &
$-\frac{ \G M(r^2-12\,r^2_\s{g})}{2\,r^3}$ \\[4pt]
$\Omega^2$ & $\frac{ \G M}{r^3}$ & $\frac{ \G M}{r^3}$ & 
$\frac{ \G M(r-2\, \rg   )^2}{r^4(r-3\,\rg )}$ & 
$\frac{ \G M}{r(r-2\, \rg   )^2}$ &
$\frac{ \G M(r^2-6\,\rg\,r +36\,r^2_\s{g})}{r^5}$\\[4pt]
$(\Omega^\s{\parallel})^2$ & $\frac{ \G M}{r^3}$ &
$\frac{ \G M(r-6\,r_\s{g})}{r^4}$ & 
$\frac{ \G M(r-6\,r_\s{g})(r-2\, \rg )^2}{r^5(r-3\,r_\s{g})}$ &
$\frac{ \G M(r-6\,r_\s{g})}{r(r-2\, \rg )^3}$ &
$\frac{ \G M(r^2-36\,r^2_\s{g})}{r^5}$\\[4pt]
\hline
\end{tabular}
\end{center}

\label{lastpage}

\end{document}